\documentclass[twocolumn,amsmath,amssymb,floatfix,prl]{revtex4-1}
\usepackage{amsmath}
\usepackage{amssymb}
\usepackage{graphicx}

\def\Rb87{^{87}\rm{Rb}}					

\begin{document}
\title{The Peierls substitution in an engineered lattice potential}

\author{K.~Jim\'{e}nez-Garc\'{\i}a$^{1,2}$}
\author{L.~J.~LeBlanc$^1$}
\author{R.~A.~Williams$^1$}
\author{M.~C.~Beeler$^1$}
\author{A.~R.~Perry$^1$}
\author{I.~B.~Spielman$^1$}
\email{ian.spielman@nist.gov}
\affiliation{$^1$Joint Quantum Institute, National Institute of Standards and Technology, and University of Maryland, Gaithersburg, Maryland, 20899, USA}
\affiliation{$^2$Departamento de F\'{\i}sica, Centro de Investigaci\'{o}n y Estudios Avanzados del Instituto Polit\'{e}cnico Nacional, M\'{e}xico D.F., 07360, M\'{e}xico}

\date{\today}

\begin{abstract}
Artificial gauge fields open new possibilities to realize quantum many-body systems with ultracold atoms, by engineering Hamiltonians usually associated with electronic systems. In the presence of a periodic potential, artificial gauge fields may bring ultracold atoms closer to the quantum Hall regime. Here, we describe a one-dimensional lattice derived purely from effective Zeeman-shifts resulting from a combination of Raman coupling and radiofrequency magnetic fields. In this lattice, the tunneling matrix element is generally complex. We control both the amplitude and the phase of this tunneling parameter, experimentally realizing the Peierls substitution for ultracold neutral atoms.
\end{abstract}

\maketitle

Ultracold atoms subjected to artificial gauge fields can realize phenomena usually in the domain of electronic systems. Prime examples include the quantum Hall effect (for abelian gauge fields), and topological insulators (for non-abelian gauge fields)~\cite{Dalibard2010}. Many of these phenomena are predicted to occur at extremely low temperatures, and adding a lattice potential to an ultracold system can increase the energy scales at which strongly correlated states are expected to emerge~\cite{Sorensen2005, Hafezi2007}. Current techniques for generating periodic potentials in ultracold atom systems use optical standing waves created with suitably polarized counterpropagating lasers~\cite{Grimm2000}.
In contrast, we describe a one-dimensional (1D) ``Zeeman lattice'' for ultracold atoms created with a combination of radiofrequency (rf) and optical-Raman coupling fields, without any optical standing waves.
In this lattice, atoms acquire a quantum mechanical phase as they hop from site to site, explicitly realizing the Peierls transformation~\cite{Hofstadter1976} in the laboratory frame.
Our approach extends existing Raman dressing schemes~\cite{Lin2009_Bfield} by simultaneously generating an artificial gauge field and an effective lattice potential.

Optical lattices generally result from the electric dipole interaction between an atom and the electric field of an optical standing wave, yielding a potential $V_{\rm{dip}}(\bf{r})\!\propto\!\alpha(\lambda)I(\bf{r})$, where $\alpha(\lambda)$ is the atomic polarizability at laser wavelength $\lambda$, and $I(\bf{r})$ is the spatial intensity distribution~\cite{Grimm2000}.
In such lattices, the natural units of momentum and energy are given by the single photon recoil momentum $\hbar k_{L}\!=\!2\pi\hbar/\lambda$ and its corresponding energy $E_{L}\!=\!\hbar^2k_{L}^2/2m$, where $m$ is the atomic mass.

Quantum particles with charge $q$ in a 1D periodic potential (here along ${\bf e}_x$) acquire a phase $\phi_j\!=\!(q/\hbar)\!\int_{x_j}^{x_{j+1}}\!{\bf A}\cdot\!{\bf e}_x{\rm d}x$ upon tunneling from site $j$ to $j+1$ in the presence of a vector potential ${\bf A}$.
For sufficiently strong potentials, this system is described by the tight-binding Hamiltonian
\begin{equation}
H=-\sum_{j}[t \exp{\!(i\phi_j)}\hat{a}^\dag_{j+1} \hat{a}_{j} + {\rm h.c.}],
\vspace{-10pt}
\end{equation}
where $\hat{a}^\dag_j$ describes the creation of a particle at site $j$, and $t\exp{\!(i\phi_j)}$ is the complex matrix element for tunneling between neighboring sites.
Using the phases $\phi_j$ to represent the effect of ${\bf A}$ is known as the Peierls substitution~\cite{Hofstadter1976}, and for uniform phase $\phi$ the energy is $E(k_x)\!=\!-2t\cos(\pi k_x/k_{L}\!-\!\phi)$, where $k_x$ is the particle's crystal momentum.
\begin{figure}
  \begin{center}
  \includegraphics[width=3.45in]{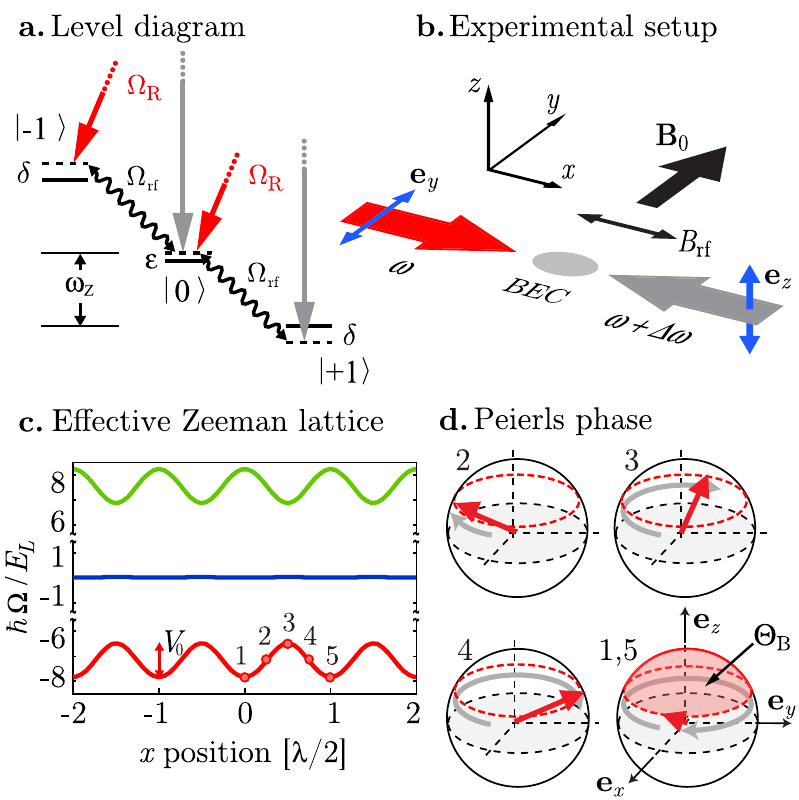}\\
  \end{center}
  \vspace{-19pt}
  \caption{Effective Zeeman lattice.~{\bf a-b.}~A uniform magnetic field $B_0 {\bf e}_y$ Zeeman-splits the levels in the $F\!=\!1$ ground state manifold of $\Rb87$ by $\omega_{\rm Z}$, and provides a quadratic Zeeman shift $\epsilon$.~In conjunction with an rf magnetic field $B_{\rm rf}{\bf e}_x$ with frequency $\Delta\omega$, a pair of orthogonally-polarized counterpropagating Raman beams with frequencies ($\omega, \omega\!+\!\Delta\omega$) illuminates the atomic sample. The rf and Raman fields have coupling strengths $\Omega_{\rm rf}$ and $\Omega_{\rm R}$.~{\bf c.}~The spatially varying eigenvalues of $\hat{H}_{\rm rf+R}(x)$ (red, blue and green curves) give rise to our $\lambda/2$ effective Zeeman lattice; as plotted $\hbar\Omega_{\rm rf}\!=\!1E_{L}$, $\hbar\Omega_{\rm R}\!=\!10E_{L}$, and $\hbar\delta\!=\!2E_{L}$.~{\bf d.}~Spatial precession of ${\bf B}_{\rm eff}(x)$ (dark arrow) and the solid angle $\Theta_{\rm B}$ it subtends when an atom tunnels to the nearest neighboring site (points 1 to 5 in {\bf c.}). This geometrical Berry's phase gives the Peierls phase $\phi$.}
  \vspace{-10pt}
  \label{Potential}
\end{figure}

We realize the Peierls substitution for ultracold atoms by synthesizing a 1D effective Zeeman lattice that allows independent control of both $t$ and $\phi$. Previous experiments (a) controlled the amplitude and sign of $t$ in driven optical lattices~\cite{Ciampini2011}, or in addition (b) controlled $\phi$ by means of rotating optical lattices~\cite{RAWilliams2010} or Raman-assisted tunneling in an optical superlattice~\cite{Aidelsburger2011}. Our effective Zeeman lattice technique provides both a periodic potential and an artificial vector potential in the laboratory frame.

The Zeeman lattice arises from a combination of rf and Raman fields that simultaneously couple the spin states $\{|{m_F}\!\rangle\}_{m_F=0,\pm1}$ of $\Rb87\mathrm{'s}$ $F\!=\!1$ ground level, which are split by $\hbar\omega_Z$ (Fig.~\ref{Potential}a,b). In the frame rotating at the rf frequency $\Delta\omega$ and under the rotating wave approximation, the combined rf-Raman coupling contributes a term
\vspace{-15pt}
 \begin{equation}\label{rfRamanHam}
    \hat{H}_{{\rm rf+R}}(x)={\bf\Omega}(x)\cdot\hat{\bf{F}} + \hat{H}_{\rm{Q}}
\end{equation}
to the overall Hamiltonian, where $\hat{\bf{F}}\!=\!(\hat{F}_x,\!\hat{F}_y,\!\hat{F}_z)$ is the $F\!=\!1$ angular momentum operator;
${\bf\Omega}\!\!=\!\!\left(\Omega_{\rm{rf}}\!+\!\Omega_{\rm R}\!\cos(2k_{L}x),\!-\Omega_{\rm{R}}\!\sin(2k_{L}x),\!\sqrt{2}\delta\right)/\sqrt{2}$, in which $\Omega_{\rm rf}$ and $\Omega_{\rm R}$ are the rf and Raman coupling strengths, and
$\delta\! = \!\Delta\omega\!-\! \omega_Z\!$~is the detuning from Raman resonance; and
$H_{\rm Q}\!\!=\!\!-\epsilon(\hbar^2 \mathbb{I}\!\! -\!\! \hat{F}_{z}^2)/\hbar$ describes the quadratic Zeeman shift.
Equation~(\ref{rfRamanHam}) is the Zeeman Hamiltonian for an effective field
${\bf{B}}_{\rm{eff}}(x)\!\!=\!\!\hbar{\bf\Omega}(x)/{g_F\mu_{\rm{B}}}$, where
$\mu_{\rm B}$~is Bohr's magneton and $g_F$~is the Land\'{e} $g$-factor. This spatially varying effective Zeeman shift produces a 1D lattice potential (Fig.~\ref{Potential}c). As atoms tunnel from site to site, ${\bf B}_{\rm eff}$ rotates by $2\pi$ about ${\bf e}_z$ (Fig.~\ref{Potential}d) and the atoms acquire a geometrical Berry's phase~\cite{Berry1984} proportional to the enclosed solid angle $\Theta_{\rm B}$.
The tunneling parameters $t$ and $\phi$, obtained from $E(k_x)$, are non-trivial functions of ${\bf \Omega}$.

When $\Omega_{\rm rf}\!\gg\!\Omega_{\rm R}, \delta, \epsilon$ the effective Zeeman shift reduces to $\hbar |{\bf \Omega}|\! \approx\! \hbar[ \Omega_{\rm rf}\! + \!\Omega_{\rm R} \cos( 2k_{L} x )]/\sqrt{2}$, and when $\Omega_{\rm R}\!\gg\!\Omega_{\rm rf}, \delta$ we obtain the analogous result $\hbar |{\bf \Omega}|\! \approx\! \hbar[ \Omega_{\rm R}\! +\! \Omega_{\rm rf} \cos( 2k_{L} x )]/\sqrt{2}$.
In both of these limits, the larger of the two fields defines a natural quantizing axis about which the smaller field spatially modulates $|{\bf \Omega}|$. For $\Omega_{\rm R}\!\gg\!\Omega_{\rm rf}$, this quantizing axis is spatially rotating.

We experimentally characterize the lattice in three ways: (i)~we measure the effective mass~$m^*\!\!=\!\!\hbar^2[{\rm d}^2 E(k_x)/{\rm d}k_x^2]^{-1}$, which in the tight-binding regime is inversely proportional to $t$; (ii)~we quantify the Peierls phase $\phi$ and test its robustness against fluctuations in $\Omega_{\rm rf}$; and (iii)~we investigate the diffraction of BECs from our effective Zeeman lattice.
In each case, we start with $\Rb87$ BECs in the $|F\!\!=\!1\!,m_F\!\!=\!\!-1\rangle$ state in a crossed optical dipole trap with frequencies $(f_x, f_y, f_z)\! =\! (13, 45, 90)$~Hz \footnote{In some experiments, both the trap frequencies and $\lambda$ were slightly different: $(f_x, f_y, f_z)\!=\!(17.3, 41.4, 90)$~Hz, and $\lambda\!=\!790.14$ nm}. In the presence of a uniform bias field $B_0 {\bf{e}}_y$, we apply an rf magnetic field with frequency $\Delta\omega/2\pi\!=g_F\mu_{\rm B} B_0=\!3.25$~MHz and prepare the BEC in the lowest energy rf-dressed state~\cite{Lin2009}.
Two $\lambda\!=\!790.33$~nm Raman laser beams, counter-propagating along ${\bf{e}}_x$ and differing in frequency by $\Delta\omega$, couple the BEC's internal degrees of freedom with strength $\Omega_{\rm {R}}$ (Fig.~\ref{Potential}a,b). The combination of rf and Raman coupling creates a 1D lattice potential along ${\bf e}_x$, the direction of momentum exchange defined by the Raman beams.
\begin{figure}
  \begin{center}
  \includegraphics[width=3.45in]{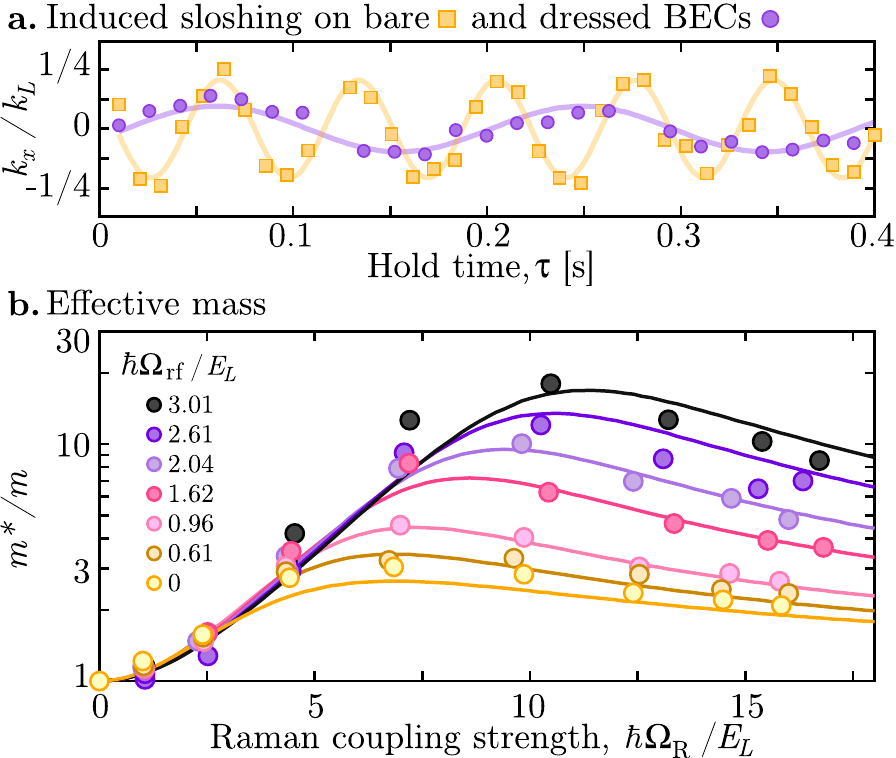}\\
  \end{center}
  \vspace{-19pt}
  \caption{Effective mass.~{\bf a.}~Comparison of the oscillations of a BEC in the $|m_F=-1\rangle$ state to those in an rf-Raman dressed BEC [$\hbar\Omega_{\rm R}\!=\!12.4(9)$~$E_{L}$ and $\hbar\Omega_{\rm rf}\!=\!2.04(6)$~$E_{L}$].~The curves are fits to a sinusoid from which we obtain $f_x\!=\! 14.0(1)$~Hz and $f^*\!=\! 5.3(1)$~Hz, thus ${m^*/m}=7.0(3)$ and $t=0.015(1)E_{L}$.~{\bf b.}~Measurements of $m^*/m$ as a function of $\Omega_{\rm R}$ and $\Omega_{\rm rf}$. The curves depict the expected $m^*/m$ ratio.}
  \vspace{-10pt}
  \label{EffMass_Slosh}
\end{figure}

We obtain the atoms' effective mass $m^*$ by inducing dipole oscillations~\cite{Lin2011_Efield,Supplemental2012,Chen2012} along ${\bf e}_x$ and measuring shifts in the oscillation frequency as a function of the coupling strengths $\Omega_{\rm R}$ and $\Omega_{\rm rf}$. The atoms slosh in the lattice for a variable time $\tau$, after which we remove all coupling and confining potentials (thus projecting the final spin-momentum superposition into bare atomic states) and absorption image after a 28.2 ms time-of-flight (TOF).
Figure~\ref{EffMass_Slosh}a shows bare and dressed condensates oscillating at frequencies $f_x$ and $f^*$, respectively; Fig.~\ref{EffMass_Slosh}b shows that the effective to bare mass ratio $m^*/m\!=\!(f_x/f^*)^2$ as a function of $\Omega_{\rm R}$ and $\Omega_{\rm rf}$ is in good agreement with calculations (curves~\cite{Supplemental2012}). These data provide the tunneling matrix element amplitude $t/E_{L}=(m/m^*)/\pi^{2}$ in the tight-binding regime~\footnote{The general relation between effective mass and tunneling amplitude is $m/m^*(k_x)\!\!=\!\!\pi^{2}\cos(\pi k_x/k_{L}\!\!-\!\!\phi)t/E_{L}$. For small sloshing amplitudes $\Delta k_x$ around the minimum $k_x\!=(\phi/\pi) k_{L}$, the effective mass is almost uniform.}.

\begin{figure}
  \begin{center}
  \includegraphics[width=3.45in]{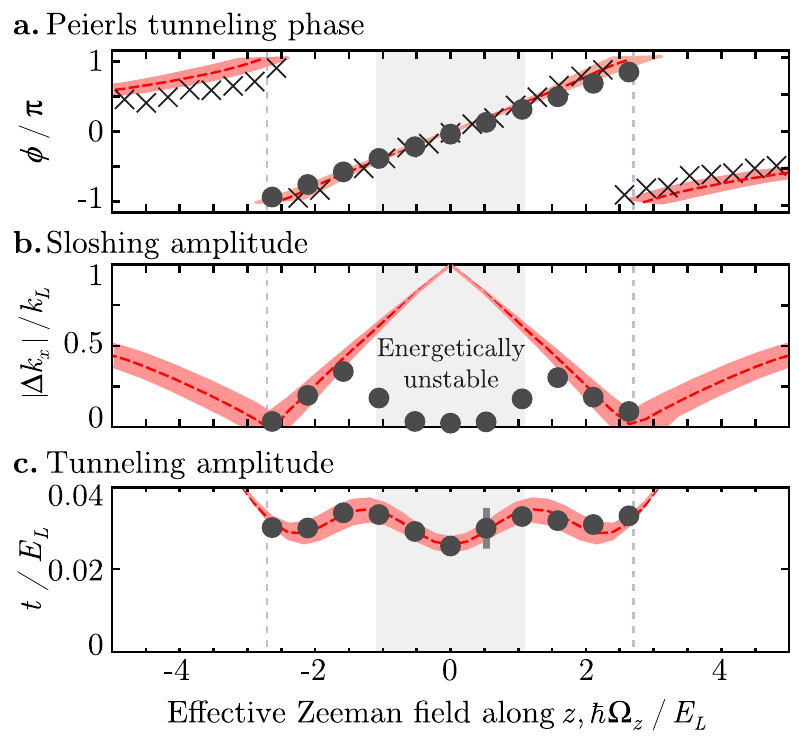}\\
  \end{center}
  \vspace{-19pt}
  \caption{Peierls transformation.~{\bf a.}~Peierls phase $\phi$ measured using adiabatic (crosses) and sudden (circles) changes of $\Omega_z$. Vertical lines denote the first Brillouin zone.~{\bf b.}~Sloshing amplitude after suddenly changing $\Omega_z$.~We observed strong damping of oscillations in the region shaded in gray.~{\bf c.}~Tunneling amplitude $t$ measured from oscillation frequency. The rf coupling was modulated as a function of $\Omega_z$ to test the robustness of the Peierls phase~$\phi$. The Raman coupling was held at $\hbar\Omega_{\rm R}\!=\!10.0(8)E_{L}$. The dashed curves correspond to the expected behavior calculated from $H_{\rm rf+R}$, and the pink bands arise from the experimental uncertainty in $\Omega_{\rm R}$.}
  \vspace{-10pt}
  \label{TunnelingPhase}
\end{figure}

\begin{figure*}
  \begin{center}
  \includegraphics[width=7.1in]{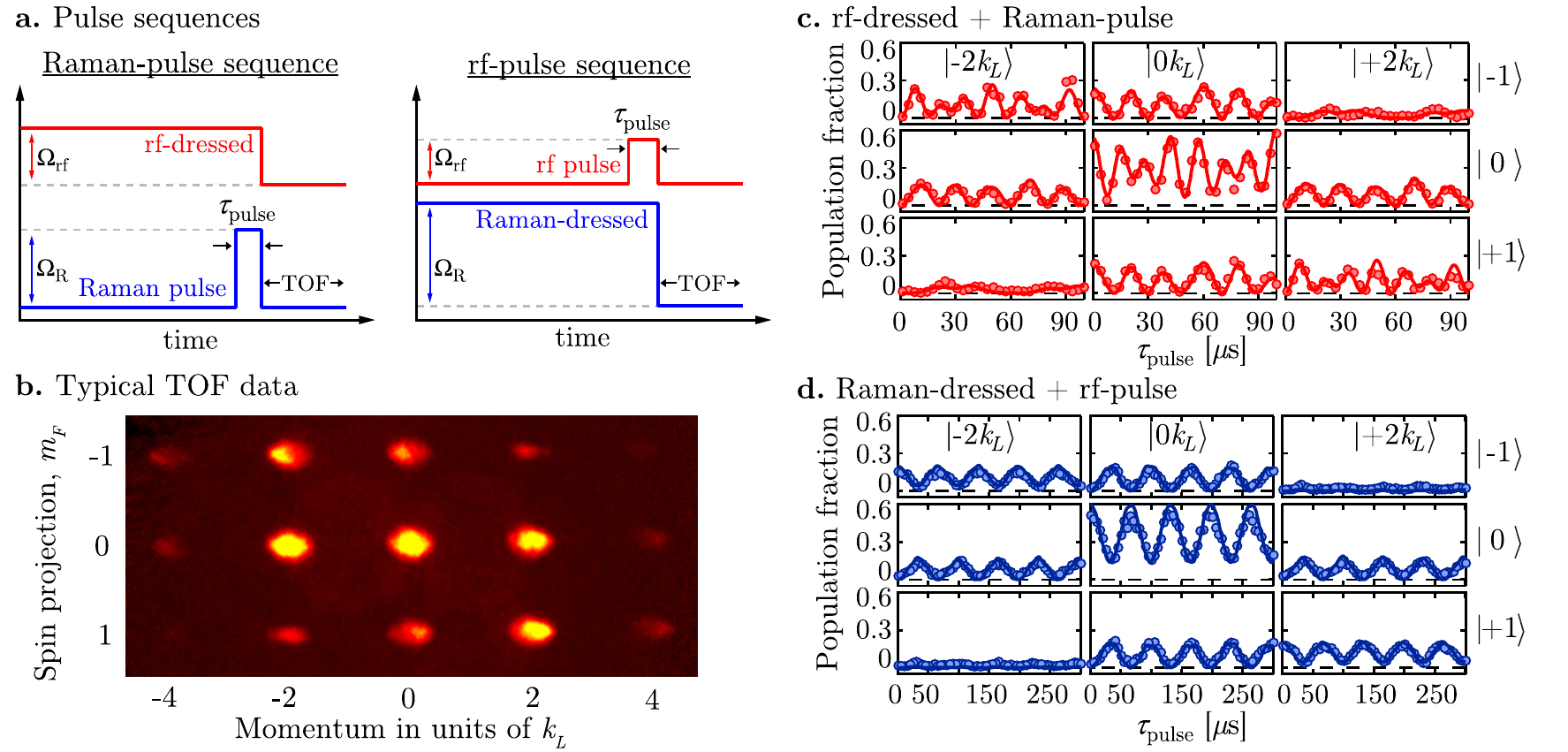}\\
  \end{center}
  \vspace{-19pt}
  \caption{BEC diffraction from the effective Zeeman lattice.~{\bf a.}~Starting with an rf-dressed (Raman-dressed) state, we suddenly turn-on the Raman (rf) field for a variable time $\tau_{\rm pulse}$.~{\bf b.}~Using TOF absorption images of the projected spin-momentum distributions, we count the number of atoms in each diffracted order and determine its fractional population. Panels~{\bf c},{\bf d} depict time evolution of these fractions. The curves are fits to the data, calculated from $H_{\rm rf+R}$. The fit parameters are:~({\bf c.})~rf-dressed $\hbar(\Omega_{\rm rf},\Omega_{\rm R},\Omega_z)\!\!=\!\!(3.57,11.49,-0.04)E_{L}$ and~({\bf d.})~Raman-dressed $\hbar(\Omega_{\rm rf},\Omega_{\rm R},\Omega_z)\!\!=\!\!(3.06,15.14,0.08)E_{L}$.}
  \vspace{-10pt}
  \label{PulsingFractions}
\end{figure*}

An important characteristic of our effective Zeeman lattice is the presence of a tunable Peierls hopping phase $\phi$, which can be revealed through its effects on $E(k_x)$ and is experimentally controlled by adjusting $\Omega_z$. We quantify $\phi$ both by adiabatically modifying the bandstructure (``adiabatic method'') and by inducing oscillations in momentum space (``sudden method'', similar to above). Furthermore, we test its robustness to variations in $\Omega_{\rm rf}$ with the latter method.

In the adiabatic method, we load a BEC at $k_x=0$ and adiabatically change $\Omega_z$, such that the BEC always sits at the minimum of $E(k_x)$ located at $k_{\rm min}$. The time scale for adiabaticity is set by the modified trapping frequency $f^*$ along the direction of the Raman beams. Once $\Omega_z$ reaches its final value, we remove the trapping potential and subsequently deload all atoms into the $|m_F\!=\!+1\rangle$ spin state while mapping the occupied crystal momentum $k_x$ to free-particle momentum~\cite{Supplemental2012}. We image this distribution after a 13.1 ms TOF, revealing $k_{\rm min}$. The Peierls phase, shown as crosses in Fig.~\ref{TunnelingPhase}a, is $\phi/\pi=k_{\rm min}/k_{L}$.

In the sudden method, we test the robustness of the Peierls phase $\phi$ by first adiabatically loading to $\phi=\pm\pi$ (the condensate sits at the edge of the Brillouin zone) and then suddenly changing both $\Omega_z$ and $\Omega_{\rm rf}$~\footnote{We use $\Omega_{\rm rf}\!=\!\Omega_{\rm rf_0}\!+\!\Delta\Omega_{\rm rf} \cos( 2 \pi  f_{\rm rf}  \Omega_z)$, where  $\hbar\Omega_{\rm rf_0}\!=\!0.75E_{L}$, $\hbar\Delta \Omega_{\rm rf}\!=\!0.23E_{L}$ and $f_{\rm rf}/\hbar\!=\!0.4E_{L}^{-1}$.} to new values (changing both $\phi$ and $t$).
This results in momentum space oscillations centered at $k_{\rm min}$. After a time $\tau$ we release the BEC, and measure as above. We fit the crystal momentum dynamics with $k_x(\tau)\!=\!k_{\rm min}\! +\! \Delta k_x \cos(2 \pi \tau f^* \!+\gamma)$, where $\Delta k_x$ is the amplitude, and $\gamma$ is an overall phase-shift whose average value is $0.9(1)\pi$ for these measurements. Figure~\ref{TunnelingPhase}a~(circles) shows the measured Peierls tunneling phase as a function of $\Omega_z$.

Measurements from the adiabatic and sudden methods are in good agreement with each other and their expected values (Fig.~\ref{TunnelingPhase}a, dashed curves), highlighting the precise experimental control offered by our rf-Raman induced effective Zeeman lattice.
This agreement also demonstrates the robustness of our engineered Hamiltonian to deliberate variations in $\Omega_{\rm rf}$ of up to~$0.25E_{L}$.

The sloshing amplitude $|\Delta k_x|$ is displayed in Fig.~\ref{TunnelingPhase}b.
For large initial $|\Delta k_x|$ (shaded region) the oscillations are strongly damped, which we attribute to energetic instabilities resulting in depletion of BEC atoms~\cite{Cristiani2004}. This is evident from the departure of the oscillation amplitude from the expected value; the experimental range of the dynamical instability, $\hbar|\Omega_z|< 1.1 E_L$ (corresponding to $\Delta k_x \gtrsim 0.6k_{L}$), is shaded in gray and is in agreement with previous observations in conventional optical lattices~\cite{Cristiani2004}.
Figure~\ref{TunnelingPhase}c displays the tunneling amplitude $t$, obtained from $f^*$.
For comparison, a sinusoidal lattice would require a depth $V_0\!\approx\!8 E_{L}$ to give similar parameters.

Having discussed the behavior of atoms in the lattice's lowest band, we now explore the full lattice by suddenly turning it on, diabatically projecting a ground state BEC into higher bands. At the beginning of such a pulse($\tau_{\rm pulse}\!\ll\! \pi \hbar/\sqrt{s}E_L$, where $s\!=\!V_0/E_L$), an ordinary periodic potential, would spatially modulate the BEC's phase~\cite{Ovchinnikov1999}; our effective Zeeman lattice induces such a modulation but in a spin-dependent manner. We focus on the $\Omega_{\rm R}\!\gg\!\Omega_{\rm rf}$ and $\Omega_{\rm R}\!\ll\!\Omega_{\rm rf}$ tight-binding regimes and investigate the spin and spatial structure of our lattice. Our data extends well beyond the short-time phase modulation regime.

In the absence of either Raman or rf coupling, there is no lattice. As indicated in Fig.~\ref{PulsingFractions}a, we use two different methods to introduce our lattice on an initial spatially uniform state: (i)~starting with an rf-dressed state (with $k_x\!=\!0$), we suddenly ($<\!1$\!~$\mu$s) turn on the Raman beams; or (ii)~starting with a Raman-dressed state~\cite{Lin2009} (a superposition of $|m_F\!=\!0, k_x\!=\!0\rangle$ and $|m_F\!=\!\pm1, k_x\mp2k_{L}\rangle$), we suddenly turn on the rf-field.

After holding the lattice on for a time $\tau_{\rm pulse}$, we suddenly turn off the rf and Raman fields, together with the confining potential. The atoms are projected onto the bare spin-momentum basis and separate in TOF in the presence of a magnetic field gradient (along ${\bf e}_y$), allowing us to resolve their spin and momentum components.

We observe detectable population in states with momenta up to $|k_x|\!\leq\!4k_{L}$ (Fig.~\ref{PulsingFractions}b). We perform such experiments for $\Omega_{\rm R}/\Omega_{\rm rf}\approx3$~and~$5$. We minimize the effects of interactions by working with small BECs ($\approx\!9\!\times\!10^4$ atoms). Figures~\ref{PulsingFractions}c,d show the fraction of atoms in each diffracted order evolving with time. We observe multiple revivals of the initial spin-momentum state and find symmetry in the population dynamics of spin-momentum states with opposite momentum {\it and} opposite spin. The curves represent fits to the populations in all spin-momentum components. The parameters from the fits are all within $10\%$ for our calibrated values, demonstrates that the spin-momentum dynamics are well described by the unitary evolution of the initial states under $H_{\rm rf+R}$~\cite{Supplemental2012}.

\begin{figure}
  \begin{center}
  \includegraphics[width=3.45in]{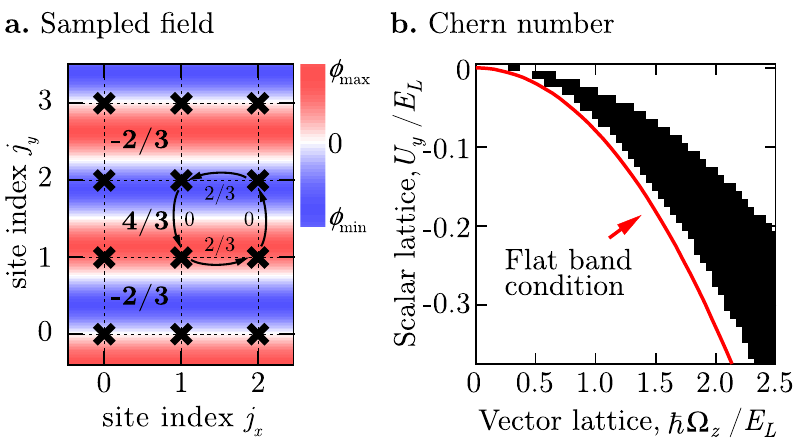}\\
  \end{center}
  \vspace{-19pt}
  \caption{$1/3$ flux Hofstadter model.~{\bf a.}~Schematic showing effective $1/3$ flux per plaquette modulo~$1$.~The color scale indicates the effective phase gradient induced by the vector lattice. Atoms acquire phase as they hop along ${\bf e}_x$, in contrast, no phase is acquired by hopping along ${\bf e}_y$ (see loop).~{\bf b.}~Region (in black) where the Chern numbers in the lowest three bands are $(1,-2,1)$, equivalent to the $n_\Phi\!=\!1/3$ Hofstadter model as a function of the period $3a/2$ vector lattice's strength. The horizontal axis is its vector contribution to $\Omega_z$ and the vertical is its scalar contribution to the overall lattice potential.}
  \vspace{-10pt}
  \label{UniformFlux}
\end{figure}

Based on this technique for controlling the Peierls phase and inspired by recent proposals for creating flux lattices~\cite{Dalibard2010,Cooper2011_PRL}, we now describe how this method might be extended to create a lattice with zero net flux that is \textit{topologically} equivalent to the Hofstadter model with flux density $n_{\Phi}=1/3$ per plaquette. Because the hopping phase is only defined modulo $2\pi$ (thus $n_{\Phi}$ is only defined modulo 1), a uniform magnetic field with $n_\Phi=1/2$ is equivalent to a staggered field with $n_\Phi=\pm1/2$. In the same spirit, a magnetic field staggered along ${\bf e}_y$ with flux density $(\ldots, 1/3, 1/3, -2/3, \ldots)$ has zero net flux yet is equivalent to a uniform field with $n_{\Phi}=1/3$. These fields could be generated by the Peierls phases~$\phi_y(j_x,j_y)\!=\!0$ and $\phi_x(j_x,j_y)\!=\!-(2\pi/3)\!\!\!\mod\!(j_y,3)$. Reminiscent of the flux rectification mechanism proposed in Ref.~\cite{Gerbier2010}, this configuration can be created in our system by adding two standing waves along ${\bf e}_y$ (normal to the Raman lasers): a state-independent lattice with period $a$ localizing the atoms to specific lattice sites (e.g.\ from a retro-reflected 532 nm laser), and a state-dependent ``vector" lattice with a period $3a/2$ sinusoidally modulating $\Omega_z$ as a function of $y$ (e.g., from an additional 790 nm laser, nearly counter-propagating). Figure \ref{UniformFlux}a shows that with a suitable relative phase between the standing waves, the fluxes along ${\bf e}_y$ repeat with the pattern $(-2/3, 4/3, -2/3)$, giving the desired flux per plaquette. To verify this heuristic interpretation, we numerically solve the 2D bandstructure (for exact parameters see~\cite{Supplemental2012}), and as shown in Fig.~\ref{UniformFlux}b, we confirm that for a wide range of parameters, the three lowest bands are described by the same Chern numbers~\cite{Fukui2005} as are those of the $n_\Phi=1/3$ Hofstadter model:~$(1, -2, 1)$.

We realized a 1D lattice potential for ultracold atoms using only rf and Raman transitions, in which the tunneling matrix element is in general complex. This work constitutes a first step towards realizing flux lattices~\cite{Cooper2011_PRL}, in which the physics of charged particles in strong magnetic fields can be simulated. The tunability of the Peierls phase achieved with our rf-Raman lattice would allow the observation of nonlinear effects of ultracold atoms in 1D periodic potentials, such as atomic density modulations with periodicity larger than the lattice spacing~\cite{Machholm2004}.

We appreciate enlightening conversations with G. Juzeliunas, and N. R. Cooper; and we thank W. D. Phillips for a careful reading of this manuscript. This work was partially supported by the ARO with funding from DARPA's OLE program and the Atomtronics-MURI; and the NSF through the JQI Physics Frontier Center. K.J.-G. thanks CONACYT; L.J.L. thanks NSERC and M.C.B. thanks the NIST-ARRA program.

\newpage

\subsection{Supplementary Material}

\subsection{INDUCING OSCILLATIONS}

Starting with the rf-dressed state described in the main text, we ramped the Raman beams from 0 to $\Omega_{\rm R}\!>\!4$ $E_{L}/\hbar$ in 70 ms. Then we ramped the rf coupling strength to an adjustable final value $\Omega_{\rm rf}$ in 2 ms, such that an effective Zeeman lattice was created. We induce sloshing by applying a synthetic electric field~\cite{Lin2011_Efield}, achieved by ramping the $z$-component of the Zeeman field to $\hbar\Omega_z\approx 2 E_{L}$ and then jumping it back to $0$ in 2~ms.

\subsection{MOMENTUM REPRESENTATION AND DIAGONALIZATION OF $H_{\rm{rf+R}}(x)$}

In order to compute the properties of our lattice potential we numerically diagonalized the combined rf-Raman Hamiltonian $H_{\rm rf+R}$ given the experimental parameters $\Omega_{\rm rf}, \Omega_{\rm R}, \Omega_z$. For simplicity we work in the momentum space representation of $H_{\rm rf+R}$. This representation offers an alternative understanding of the effective Zeeman lattice structure that arises from the combination of rf and Raman coupling fields.

While Raman transitions couple together states with $m_F$ differing by $\pm1$ and momentum differing by $\pm2\hbar k_{L}$, rf-coupling processes only change $m_F$ by $\pm1$, leaving the momentum unchanged. The combination of both Raman and rf fields sets appropriate conditions for secondary rf and Raman processes to populate states with higher momentum. The available states under rf-Raman coupling constitute a set of spin-momentum states $\{|\Psi_n\rangle\} =\{|m_F,\hbar (k_x \!+\! 2nk_L)\rangle \}$ where $n\!\in\!\mathbb{Z}; m_F\!=\!0,\pm1$. As expected, this basis is that of a lattice.

In the basis $\{|\Psi_n\rangle\}$, $H_{\rm rf+R}$ is a Hermitian block matrix of size $3(2N\!+\!1)$, when $n$ is restricted to $n\leq N$. For our parameters, dimensions larger than $3(2N\!+\!1)\!=81$ provided indistinguishable results. The dimension of this basis is appropriate for our calculations since we observed the population of states with up to $|n|\!=\!4$. For clarity, we construct the Hamiltonian by arranging the spin-momentum states in the $\{\Psi_n\}$ basis with increasing momentum index $n\!=\!-N,\! -N\!+\!1, \ldots,N$.
Along the principal diagonal, we have $3\times 3$ blocks
\begin{equation*}\label{rfRaman_Ham_diagonal}
    \mathbb{A}_{k_x}(n)\!=\!\hbar^2(k_x\!+\!2nk_L)^2 \mathbb{I} + \frac{\Omega_{\rm rf}}{\sqrt{2}}{F}_{x}-[\Omega_z {F}_{z}+\frac{\epsilon}{\hbar} (\hbar^2 \mathbb{I} \!-\! {F}_z^2)];
\end{equation*}
these terms correspond to kinetic energy, rf coupling of spin states with equal momentum, and the real Zeeman interaction, respectively; $\{F_x, F_y, F_z\}$ are the matrix representations of the $F\!=\!1$ angular momentum operators, and $\mathbb{I}$ is the $3\!\times\!3$ identity matrix.
Above and below the $\mathbb{A}_{k_x}(n)$ blocks, we have $3\!\times\!3$ blocks $\mathbb{B}\!=\!\sqrt{2}\Omega_{\rm R}({F}_{x}\!-\!i{F}_{y})/4$ describing the Raman coupling of spin-momentum states differing in momentum by $\Delta k_x\!=\!\pm 2k_{L}$.

The diagonalization of $H_{\rm rf+R}(n)$ as a function of $k_x$ gives the bandstructure of the combined rf-Raman lattice potential, $E(k_x)\!\!=\!\! E(k_x, \Omega_{\rm rf}, \Omega_{\rm R}, \Omega_z, \phi)$.

\subsection{EFFECTIVE ZEEMAN LATTICE PROPERTIES}

We extract the Zeeman lattice properties by fully characterizing the lowest energy band which in the tight binding regime is of the form $E(k_x)\!\!=\!\!-2t\cos(\pi k_x/k_{L}\!\!-\!\!\phi)$.
The tunneling amplitude is given by $t=\Delta E/4$, where $\Delta E$ is the width of the lowest band.
The lattice depth was calculated as the depth that a lattice potential would have in order to produce the calculated width $\Delta E$.
The effective mass is defined as $m^*\!=\!\hbar^2 [{\rm d}^2 E(k_x)/{\rm d}k_x^2]^{-1}$, where the derivative is evaluated at the point of interest (e.g. for our $\Omega_z=0$ measurements of effective mass displayed in Fig.~2, this was at $k_x=0$).
In the tight-binding regime, $t$ and $m^*/m$ are inversely proportional to each other $\pi^2 t/E_{L} \!=\! m/m^*$.
We obtained the Peierls phase $\phi$ by computing the shifts the lowest band [with minimum at  $k_{\min}\!= \!(\phi/\pi) k_{L}$] as a function of the experimental parameters.

\subsection{DELOADING}

We deloaded to a single bare spin state by rapidly ramping to $\Omega_z\!=\!0$, taking $\hbar\Omega_{\rm R}\!\rightarrow\!0$ to zero in 500~$\mu$s while increasing $\hbar\Omega_{\rm rf} \!\rightarrow\! 3E_{L}$. We then ramped the $\hbar\Omega_z\!\rightarrow\! -140E_{L}$ transferring all atoms into $|m_F\!=\!+1\rangle$.

\subsection{2D LATTICE FOR $1/3$ FLUX\\ HOFSTADTER MODEL}

We numerically solve the bandstructure of the 2D lattice arising from the combination of a period~$a\!=\!266$~nm lattice with depth $V_0=22.5 E_L$, and a~1D ``Zeeman lattice" with $\Omega_{\rm rf}\!=\!1 E_L$, $\Omega_{\rm R}\!=\!10 E_L$, $\epsilon\!=\! 0.44 E_L$. Here~$E_L$~denotes the Raman recoil, not the short-period lattice~recoil.
\end{document}